\documentclass[sigconf,screen,9pt,authorversion]{acmart}

\usepackage{booktabs}

\usepackage{tikz}
\usepackage{float}
\usetikzlibrary{shapes,arrows}
\usepackage{graphicx}
\graphicspath{{}}
\usepackage{localmacros}

\makeatletter 
\def\arcr{\@arraycr}
\makeatother

\setcopyright{acmlicensed}
\acmPrice{15.00}
\acmDOI{10.1145/3283812.3283822}
\acmYear{2018}
\copyrightyear{2018}
\acmISBN{978-1-4503-6055-5/18/11}
\acmConference[NL4SE '18]{Proceedings of the 4th  ACM SIGSOFT International Workshop on NLP for Software Engineering}{November 4, 2018}{Lake Buena Vista, FL, USA}
\acmBooktitle{Proceedings of the 4th  ACM SIGSOFT International Workshop on NLP for Software Engineering (NL4SE '18), November 4, 2018, Lake Buena Vista, FL, USA}

\begin{document}
\title{Generating Comments From Source Code with CCGs}

\author{Sergey Matskevich}
\affiliation{
  \institution{Drexel University}
  \city{Philadelphia}
  \state{Pennsylvania}
  \country{USA}
}
\email{sm3372@drexel.edu}

\author{Colin S.\ Gordon}
\orcid{0000-0002-9012-4490}
\affiliation{
  \institution{Drexel University}
  \city{Philadelphia}
  \state{Pennsylvania}
  \country{USA}
}
\email{csgordon@drexel.edu}

\begin{abstract}
Good comments help developers understand software faster and provide better maintenance. However,
comments are often missing, generally inaccurate, or out of date.
Many of these problems can be avoided by automatic comment generation. This paper presents a method to generate informative comments directly from the source code using general-purpose techniques from natural language processing.
We generate comments using an existing natural language model that couples words with their individual logical meaning and grammar rules, allowing comment generation to proceed by search from declarative descriptions of program text.
We evaluate our algorithm on several classic algorithms implemented in Python.
\end{abstract}

\ccsdesc[500]{Computing methodologies~Natural language generation}
\ccsdesc[500]{Software and its engineering~Documentation}

\keywords{Comment Generation, Natural Language, Combinatory Categorial Grammar}

\maketitle

\section{Introduction}

Comments play a major role in the software development process and are an important tool for software maintenance \cite{Aggarwal:maint}. However, comments are often inaccurate, outdated, or do not exist \cite{deSouza:SDE}, making the maintenance more difficult and time-consuming. Automated comment generation systems can improve the situation by producing new, current comments.
An essential aspect of this is often describing not only what the code does, but how. 
Prior work on comment generation does not address these challenges in their full generality.
Some work generates helpful and precise comments~\cite{Sridhara:2010,Sridhara:2011,Sridhara:2012,pollock:17}
by using function and variable names for describing the actions performed~\cite{Hill:2010:Diss}.  Thus their approach would encounter problems with code that does not follow specific function and class naming rules. 
On the other hand, Wong et al. \cite{wong:miningComments} propose to mine web resources StackOverflow for collecting comments on them and using those comments as a basis for code comments. This approach can produce accurate comments, but it depends on the quality of the original posts as well as having a specific code segment available in the database. If the accepted answer on StackOverflow is wrong, the produced comment will also be wrong.
More recent work~\cite{louis2018deep} uses deep learning to associate comments with code, but the approach is probabilistic, may produce grammatically incorrect sentences, and is limited to high-level comments.  Generating comments that explain algorithms requires working with a more complete semantic model of the code, and this can be done in a way that improves the linguistic quality of the comments as well.

We propose an approach to generate comments from the source that relies only on the source itself and the information which can be extracted from the source, such as invariants or semantics.
Our approach is not dependent on function names, requires no external sources, and is guaranteed to produce comments representing the behavior of the source code.
Our approach adapts existing state of the art NLP algorithms for connecting program text, meaning, and grammatically-correct natural language.
This both allows the approach to benefit from ongoing advances in traditional NLP, and permits us to focus on deriving logical forms --- logical descriptions of the code, used as the basis for fully-general natural language generation.
We derive the logical forms directly from the source code and use an AI planner for Combinatory Categorial Grammars (CCGs)~\cite{Geib:2009,Geib:2015} to generate complete sentences from these specifications.
By using established NLP algorithms for text generation, we obtain a modular comment generation framework that is extensible to other programming languages or natural languages, and can easily incorporate new sources of information such as types or invariants inferred by other analyses.

This paper describes a systematic approach to automatic comment generation that does not rely on external information sources, and can be applied readily to natural languages beyond English.
We use NLP and AI planning techniques to generate comments from a general and extensible logical description of the code, which are not only accurate but guaranteed to be grammatically correct.
We also report early results on a functioning prototype using these approaches to generate statement-level comments for both handcrafted examples of common Python idioms, and individual open source Python implementations of common algorithms.

\section{Background}
This section describes relevant background on natural language generation and the particular formalism we use, Combinatory Categorial Grammar.

\subsection{Natural Language Generation}
Natural Language Generation is defined as a process of producing natural language output from non-linguistic input \cite{Jurafsky:2000}.
The most important elements are a \emph{discourse planner} and a \emph{surface realizer}.

Discourse planners are AI planners that use a special planning domain. Classical AI planners take a desired outcome, and generate a sequence of actions to achieve that outcome. The planning problem is specified by a planning domain, which contains all of the actions that can be taken, as well as transition rules between actions and the changes to the world state (a set of true logical predicates) after an action is taken. 
Discourse planners do this using a planning domain designed specifically for natural language generation.

In the context of the comment generation the world state is a set of predicates about the program source code, and the goal outcome is a state where the relevant facts about the code have been communicated. For example, given an assignment statement, the logical predicate would identify it as an assignment and record what variable was assigned which value. If this predicate is given to the discourse planner, it will create a discourse plan for expressing an assignment statement in English. This includes making choices about the tense, voice (active vs.\ passive), and case for the language to be generated.

Surface realizers receive fully specified discourse plans as input, and produce appropriate English sentences based on a formal language grammar~\cite{Jurafsky:2000}. 

A common way to generate text is the following. First, a communicative goal is determined --- the factual content of the sentence to generate. Then, a planner generates a sentence specification using the provided knowledge base and the goal. The planning domain contains the knowledge which allows making decisions about the generated sentence. The planner selects appropriate content and lexical items from the knowledge base for expressing the goal, which results in a set of facts and narrative choices (voice, etc.) to constrain what and how to communicate. These selections are then passed as specifications to the surface realizer which produces a natural language sentence using the provided grammar.

\subsection{Combinatory Categorial Grammar}
\label{sec:ccgs}
Combinatory Categorial Grammar (CCG)~\cite{Steedman:2001} is a multi-modal lexicalized tree grammar used in the parsing of natural language.  Its expressivity is between context-sensitive and context-free grammars (in fact, equivalent to linear indexed grammars~\cite{VijayEq:94,Kuhlmann2015}), and can express most structure of most natural languages~\cite{steedman:SyntProc,hockenmaier2006creating,ambati2018hindi}. The grammar consists of multiple \textit{categories} or \textit{types}, which can be basic or complex. The basic categories are various parts of speech, such as a noun, verb, adjective, etc. The complex categories are compound categories for entities like noun phrases or verb phrases.
CCGs include two kinds of directed function types for indicating how to combine adjacent word sequences, closely related to those in variants of ordered linear logic~\cite{lambek1958mathematics}.
The key connectives are $A/B$ --- something which, when combined with a $B$ to its right, constitutes an $A$ --- and $A\backslash B$, which is the same, but expecting an argument to the left. The direction of the slash indicates which side the argument should come from.\footnote{Note that compared to residuation-based logics like Lambek calculus~\cite{lambek1958mathematics}, CCGs reverse the domain and codomain in $A\backslash B$; in CCGs codomains are always on the left.}
In addition to those applications of the directed function types, there are other combinators and forms of subtyping, closely related to the combinatory logic~\cite{curry:58}.

\begin{figure}
	\[\begin{array}{rlrl}
	sort \Rightarrow & \lambda x.sort'(x) &:& VP/NP \\
	array \Rightarrow & \lambda y.array'(y) &:&  NP
	\end{array}\]
	\vspace{-1em}
	\caption{Example of a CCG grammar semantics}
	\label{CCG:ex1}
\end{figure}
A key aspect of using a grammar formalism in natural language understanding, machine translation, or natural language generation, is a connection between the words included in the grammar and the logical form used to represent the logical meaning of natural language. 
Lexicalized grammars like CCGs --- those where types are given to individual words, rather than phrases --- offer relatively simple connection between the syntactic words and logical meaning, since each word essentially corresponds to either a predicate on entities (in the case of nouns) or a function from predicates to predicates (in the case of verb phrases, etc.), taking the same number of arguments in the logical form as in the grammar.
Figure \ref{CCG:ex1} gives an example of a simple CCG lexicon that can parse and give semantics for the phrase \textit{sort the array}. The lambda expression is the semantics of the rule on the right hand side. In this case the grammar states that the word \textit{sort} produces a verb phrase ($VP$) and it expects a noun phrase ($NP$) as an argument to its right. The word \textit{array} produces a noun phrase that does not expect arguments. 
The lambda expressions inside the rules describe the meaning of each word --- its semantics, called logical forms. When words and phrases are combined, so are their logical forms. The determiner \textit{the} is a combinator that does not change the semantics of the phrase and therefore does not appear in the logical form. The logical form of the phrase ``sort the array'' combines Figure \ref{CCG:ex1}'s word semantics by passing the semantics of ``array'' to the semantics of ``sort'': $\lambda x.sort'(x) \hspace{3px} \lambda y.array'(y) \Rightarrow sort'(array')$.

\section{The Architecture}
\label{architecture}

\tikzstyle{block} = [rectangle, draw, fill=white!20, 
text width=3.5cm, text centered, rounded corners, minimum height=3em]
\tikzstyle{line} = [draw, -latex']

\begin{figure}[t]
	\includegraphics[width=0.8\columnwidth]{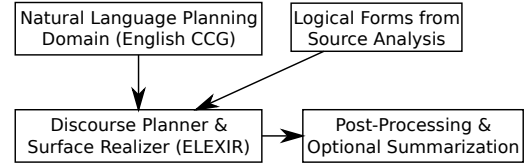}
	\caption{Comment generation pipeline}
	\label{fig:pipeline}
\end{figure}

\figref{pipeline} depicts our process for comment generation system from source. In the first step, the (official) Python parser reads the source code producing an AST, which is then translated into a logical form describing the details and structure of the source code, including program order. Each statement of the programming language represents an action or a meaning. During this step we capture this action as a set of logical forms. For example, consider an assignment statement $x = 5$. This statement will produce a logical form \textit{assign(x,5)}, which will allow us to express the fact of the assignment in English. In case of simple statements, such as an assignment, a single logical form might be enough to produce the description of it. However, we will produce several logical forms for more complex expressions such as loops or if-statements, as described in Table \ref{tbl:examples}. 
We generate logical forms for basic expressions, and combine these to produce the logical forms of more complex code structures. These logical forms are then used as the communicative goal, the input given to the AI planner in the next step.

The core of our system is an AI planner called \sysname{}~\cite{Geib:2009, Geib:2015}, which uses an extension to the basic CCG variant described in Section \ref{sec:ccgs} for the description of planning domains.  Because \sysname{} uses a \emph{grammar} formalism for the definition of a planning domain\footnote{Typically, these systems use some form of action logic or dynamic logic.}, it allows us to combine sentence planning and surface realization tasks into a single step, leading to a simpler design than with other planners~\cite{Geib:2009}.
In this case, the communicative goal given to the planner/realizer is exactly the logical form of the code, which is interpreted as a goal of giving an English description of the code represented by this logical form. 

Because  \sysname{} uses CCGs to represent planning domains, it is a straightforward process to map logical forms from the source to the syntactic structures of English (see Section \ref{sec:ccgs}). The \sysname{} grammar contains constraints in each production rule, which can be fulfilled by the given logical forms. With these constraints the \sysname{} performs A${}^*$ search within the  domain and produces a grammatically correct comment satisfying our communicative goal. 

This sentence then requires modest post-processing to transform the ``plan'' into a readable string (capitalization, removing metadata about tenses, etc.).
If the generated comments are too verbose, an off-the-shelf language summarization system could be applied (currently not implemented) to create a shorter version of the comment containing the same information.

A key benefit of this architecture is modularity. Each step in the pipeline can be improved independently of one another to improve overall system performance and quality, or to target different programming languages or natural languages. 
Replacing our Python source analysis with one for Java would allow reuse of the remaining components for Java comment generation.
Replacing the English CCG with the CCG for another language (e.g., German~\cite{hockenmaier2006creating} or Hindi~\cite{ambati2018hindi}), plus a bridge to map logical forms into the new language (as common in machine translation~\cite{phillips1993generation}) would allow reusing other components for a different natural language.
And because most of the components (planning and grammar domains, planning and realization algorithms, and summarization) were originally developed for a variety of other useful contexts, we will eventually benefit from ongoing improvements in the underlying systems.

\section{Early Results}
\begin{table*}[th!]
	\caption{Example outputs from CCG-based comment generation.}
	\label{tbl:examples}
	\vspace{-1em}
	\begin{tabular}{|p{0.25\textwidth}|p{0.25\textwidth}|p{0.45\textwidth}|}
		\hline
		Code & Logical Forms & Generated Comment\\
		\hline
		\hline
		\texttt{if x != y:} & \textit{condition(), inequality(x, y)} & Checking for inequality between x and y\\
		\hline
		\texttt{a = [\dots] for e in a:} & \textit{iterate(), element(), list(a)} & Iterate over elements of the list a\\
		\hline
		\texttt{a = \{\dots\} for e in a:} & \textit{iterate(), keys(), dictionary(a)} & Iterate over the keys of the dictionary a\\
		\hline
	\end{tabular}
	\vspace{-1em}
\end{table*}

To evaluate our algorithm we have implemented the pipeline from Figure \ref{fig:pipeline} for a subset of Python complete enough to implement classic algorithms, and have run it on both a set of manually-crafted set of code snippets, and the source for some example Python programs. The current subset of Python we cover includes loops, if-statements, IO operations, assignments, boolean expressions, function definitions, function calls built-in Python data structures. We do not, however, include syntactic sugar, such as list comprehensions. There are two reasons why we start with the basic code structures. One is the use of the algorithm for educational purposes. Two, logical forms for basic expressions can be combined to describe more complicated code fragments.

The manually-created code fragments were hand-picked to exemplify a range of common Python idioms, such as iteration over lists vs.\ dictionaries, assignments of different types of expressions, and other points of contrast.
For the second test we selected a set of the following well-known algorithms from an online repository\footnote{https://github.com/TheAlgorithms/Python}: md5 hash calculation, finding the longest subsequence, bubble sort, shell sort, radix sort, Simpson rule, and trapezoidal rule. These algorithms were selected because together they provide good coverage of the currently supported capabilities of our prototype.

We hand-crafted an English grammar for the \sysname{} that includes a minimal subset of English which can describe all code structures for our chosen subset of Python. Currently, our system can only produce a description for one statement at a time, and not a group of related statements, but is complete enough to validate the feasibility of our general and modular approach. We are currently also limited by the size of our lexicon, which prevents us from generating multiple variations of sentences, and limits us to a fixed set of ``known'' variable names. We currently support only the identifiers required for the subject code so far, but general solutions to this are known for CCGs. 

Table \ref{tbl:examples} gives a few examples of code snippets, the logical forms generated from the parse tree, and the comments synthesized from those forms using \sysname{}.  There are a couple of things to notice about the examples, beyond that they are grammatically correct and describe the relevant code.  First, the language choices, while constrained by our lexicon, involved no kind of template --- the only inputs to the generation were the logical forms and the English lexicon giving grammar and logical meanings for words. 
Second, while the second and third rows include both a definition and use of \texttt{a}, the information about \texttt{a} being a list or dictionary need not come from the parse tree, but in general could come from another program analysis such as type- or invariant-inference --- an eventual goal supported by the use of logical forms, which allow for adding other sources of relevant information.

In total we have run our prototype on 121 examples: 20 manually-created and 101 lines from TheAlgorithms (out of 245; the omitted lines are function declarations or uses of currently-unsupported features like list comprehensions).  Manual inspection of the logical forms and generated English for all manual examples and 101 of the lines of library code suggest this approach is capable of generating comments with high language quality, and factual accuracy.

\section{Related Work}

Most related work on comment generation is by Sridhara et al.~\cite{Sridhara:2010, Sridhara:2011, Sridhara:2012, pollock:17, Pollock:2013}, with important extensions by McBurney and McMillan~\cite{McBurney:2014}. They use techniques adapted from textual code search work, specifically the Software Word Usage Model (SWUM)~\cite{Hill:2010:Diss}.
SWUM models the relationships between objects (in the object-oriented sense) in the source code, in particular action relationships (such as one object invoking another's method). The model is built using statistical NLP techniques~\cite{manning99foundations}, and maps verbs and phrases that appear in function names to the expression of their functionality in natural language. The model performs well in terms of comment accuracy, but reliance on the names of functions and identifiers makes it unstable. If the naming convention of the source is different from the assumptions made by SWUM, it will not work well. In addition, the model does not focus on the quality of the sentences, so it is possible for it to generate grammatically incorrect sentences that are difficult to read. Sridhara et al.'s SWUM-based work generates comments for the abstractions and their relationships in the code. McBurney and McMillan~\cite{McBurney:2014} extend their work to generate summaries of the context where the methods are used. Their work inherits the disadvantages of the SWUM model. 

Wong et al. \cite{wong:miningComments} mine web resources such as StackOverflow to generate comments. This can work well in principle, but it has two major drawbacks: it requires that a similar code fragment to be posted on the site, and that the accepted response is correct. There is no way to automatically verify correctness of the best answer, so it is possible for this model to produce incorrect comments.

Louis et al.~\cite{louis2018deep} use deep learning to recognize redundant comments.  A byproduct of their approach is a statistical model associating code with likely comments, which can be sampled to generate likely comments.  This can generate useful high level comments.  However, the comments may be grammatically or factually incorrect because the approach lacks both a language model and a ground truth relationship between the code and language.

\section{Future Work}
The results reported are for the early stages of our work, with several clear avenues for improvement.
A key long-term project is crafting a comprehensive planning domain for generating comments. \sysname{} uses a different flavor of CCG from OpenCCG\footnote{https://github.com/OpenCCG/openccg} , and therefore lacks features that would permit reusing or trivially translating its existing grammars.
We can also explore automatic construction our planning domain, either via an automated converter from OpenCCG grammars to \sysname{} grammars (non-trivial due to the different CCG variants), or using an automated grammar learner that constructs an English grammar targeted for describing programs, trained on a large body existing comments. 
Another key step is to expand our algorithm to work on a set of related statements. In many cases, a set of statements perform a single action, for example initializing a complex object can take several statements. At the moment our algorithm cannot generate comments for such cases.

Our ultimate goal is to be able to describe known algorithms based only on program semantics, such as recognizing an algorithm as some kind of sort.
This is one place having a logical representation can simplify further work: recognizing algorithms is a form of logical consequence.
 
\bibliographystyle{ACM-Reference-Format}

\end{document}